\documentclass[aps,pre,twocolumn,superscriptaddress,floatfix]{revtex4-1}
\usepackage{graphicx}
\usepackage{amsmath}
\usepackage{amssymb}
\usepackage{color}
\newcommand{\revis}[1]{#1}
\begin{document}

\title{Thermodynamics of polymer adsorption to a flexible membrane}
\author{Steffen Karalus}
\email[E-mail: ]{karalus@thp.uni-koeln.de}
\affiliation{Institut f\"ur Theoretische Physik,
Universit\"at Leipzig, Postfach 100\,920, D-04009 Leipzig,\\
and Centre for Theoretical Sciences (NTZ), Emil-Fuchs-Stra{\ss}e 1, D-04105 Leipzig, Germany}
\affiliation{Institut f\"ur Festk\"orperforschung (IFF-2) 
and Institute for Advanced Simulation (IAS-2), 
Forschungszentrum J\"ulich, D-52425 J\"ulich, Germany}
\affiliation{Institut f\"ur Theoretische Physik, Universit\"at zu K\"oln,
  Z\"ulpicher Stra{\ss}e 77, D-50937 K\"oln, Germany}
\author{Wolfhard Janke}
\email[E-mail: ]{Wolfhard.Janke@itp.uni-leipzig.de}
\homepage[\\ Homepage: ]{http://www.physik.uni-leipzig.de/CQT.html}
\affiliation{Institut f\"ur Theoretische Physik,
Universit\"at Leipzig, Postfach 100\,920, D-04009 Leipzig,\\
and Centre for Theoretical Sciences (NTZ), Emil-Fuchs-Stra{\ss}e 1, D-04105 Leipzig, Germany}
\author{Michael Bachmann}
\email[E-mail: ]{Bachmann@smsyslab.org}
\homepage[\\ Homepage: ]{http://www.smsyslab.org}
\affiliation{Institut f\"ur Festk\"orperforschung (IFF-2) 
and Institute for Advanced Simulation (IAS-2), 
Forschungszentrum J\"ulich, D-52425 J\"ulich, Germany}
\affiliation{Center for Simulational Physics, The University of Georgia, Athens, Georgia 30602, USA}
\begin{abstract}
We analyze the structural behavior of a single polymer chain grafted to an attractive, flexible surface.  Our model is composed of a coarse-grained bead-and-spring polymer and a tethered membrane.  By means of extensive parallel tempering Monte Carlo simulations it is shown that the system exhibits a rich phase behavior ranging from highly ordered, compact to extended random coil structures and from desorbed to completely adsorbed or even partially embedded conformations.  These findings are summarized in a pseudophase diagram indicating the predominant class of conformations as a function of the external parameters temperature and polymer-membrane interaction strength.  By comparison with adsorption to a stiff membrane surface it is shown that the flexibility of the membrane gives rise to qualitatively new behavior such as stretching of adsorbed conformations.
\end{abstract}
\maketitle
\section{\label{sec:intro}Introduction}
The interaction of macromolecules and cell membranes is essential for almost all
biological processes.  Membrane proteins like glycoproteins and transmembrane
proteins govern the exchange of signals, small molecules, and ions between the
intracellular and extracellular solvent.  Membrane-embedded receptors are specific for
the binding of ligands.  The conformational changes caused by the binding process
can trigger cellular motion, drug delivery, or enzymatic catalysis.

\revis{It is interesting to understand how the
  conformational changes that a polymer can experience in the binding process to a
  membrane-like substrate depend on external parameters such as temperature and
  adsorption strength. Our study aims at the systematic investigation of such transitions
  for classes of polymer--membrane systems. Therefore, it is our goal to construct a
  conformational phase diagram that comprises the generic phase behavior of these
types of organic hybrid systems.}

Initiated by a different motivation, the study of binding affinity and specificity of
organic and inorganic matter, much work has been dedicated to the identification
of structural transitions polymers experience when adsorbing to solid
substrates \cite{hegger_chain_1994, vrbova_adsorption_1996, vrbova_adsorption_1998, vrbova_adsorption_1999, singh_crossover_2001, rajesh_adsorption_2002, krawczyk_layering_2005, bachmann_conformational_2005, bachmann_substrate_2006, kallrot_dynamic_2007, moeddel_conformational_2009, moeddel_systematic_2010}.  In these studies, the substrate is typically considered as a
solid material (e.g., a crystal) with virtually no thermal activity (i.e., its
surface structure does not change in the course of thermal fluctuations).

Although being equally important, much less is known about the thermodynamic structural
behavior of a polymer that interacts with a fluctuating surface such as a
membrane \cite{breidenich_shape_2000, breidenich_adsorption_2001,
  auth_self-avoiding_2003, auth_fluctuation_2005}.  

In this paper, we are going to study a simple coarse-grained model system
consisting of a flexible, elastic polymer grafted to a fluctuating substrate by
means of generalized-ensemble Monte Carlo computer simulations.  In a comparative
analysis, we show how the structural phase diagram changes under the influence of
thermal membrane fluctuations.

The paper is organized as follows.  In Sec.~\ref{sec:modelmethods} the model system is described in detail.  Also, the parallel tempering Monte Carlo method is reviewed shortly and the measured observables are introduced.  Section~\ref{sec:results} presents and discusses the main results, the pseudophase diagrams for the two systems under comparison.  Finally, Sec.~\ref{sec:summary} concludes the paper with a summary of our findings.

\section{\label{sec:modelmethods}Model and methods}
\subsection{\label{sec:model}Compound polymer--membrane model}
We employ a coarse-grained off-lattice model for \revis{a single elastic flexible
  homopolymer~\cite{grest_molecular_1986, schnabel_surface_2009,
    schnabel_elastic_2009, seaton_developments_2008, seaton_wang_2009}
  consisting of $N = 13$ monomers \cite{rem1}}. All monomers interact pairwise via a Lennard-Jones (LJ) potential,
\begin{equation}
  \label{eq:V_LJ_pp}
  V_\mathrm{LJ}^\mathrm{pp} (r) =
  4 \epsilon_\mathrm{pp} \left[ \left( \sigma/r \right)^{12} - \left( \sigma/r \right)^6 \right] \,,
\end{equation}
modeling van der Waals forces.  Here, $r$ denotes the relative distance between the monomers, $\sigma = r_0 / 2^{1/6}$ the zero of the LJ potential, and $r_0$ the minimum-potential distance.  The interaction strength is determined by the energy parameter $\epsilon_\mathrm{pp}$.  Throughout this study we set $r_0 \equiv 1$ and $\epsilon_\mathrm{pp} \equiv 1$ as basic length and energy scales.
Additionally, adjacent monomers are tied together by the finitely extensible nonlinear elastic (FENE) potential~\cite{bird_dynamics_1987,milchev_formation_2001},
\begin{equation}
  \label{eq:V_FENE_p}
  V_\mathrm{FENE}^\mathrm{p} (r) =
  - \frac{K}{2} R_\mathrm{p}^2 \ln\left\{ 1 - \left[ (r - r_0) / R_\mathrm{p} \right]^2 \right\}\,,
\end{equation}
modeling covalent bonds.  
$r_0$ is the same minimum-potential distance as above.  The FENE potential
behaves like a harmonic potential with spring constant $K$ in the vicinity of
$r_0$ and diverges for $r \rightarrow r_0 \pm R_\mathrm{p}$.  For the
simulations we set $K \equiv 40$ and $R_\mathrm{p} \equiv 0.3$.
The conformation of the polymer is completely defined by the set of position vectors $\{\mathbf{r}_i\}_{i = 1, \dots, N}$.  Eventually, the contribution of the polymer to the total energy reads
\begin{equation}
  \label{eq:pol_energy}
  E^\mathrm{pol} =
  \sum_{i=1}^{N-1} \sum_{j=i+1}^{N} V_\mathrm{LJ}^\mathrm{pp} ( \left| \mathbf{r}_i - \mathbf{r}_j \right| )
  + \sum_{i=1}^{N-1} V_\mathrm{FENE}^\mathrm{p} ( \left| \mathbf{r}_i - \mathbf{r}_{i+1} \right| )\,.
\end{equation}

The fluctuating substrate is modeled by a tethered membrane~\cite{kantor_statistical_1986,popova_structure_2007,popova_adsorption_2008} where the individual building segments (nodes) are tied together according to a square lattice structure with $L_x \times L_y$ nodes in total.  As tethering potential we apply, again, a FENE potential,
\begin{equation}
  \label{eq:V_FENE_m}
  V_\mathrm{FENE}^\mathrm{m} (r) =
  - \frac{K}{2} R_\mathrm{m}^2 \ln\left\{ 1 - \left[ (r - r_0) / R_\mathrm{m} \right]^2 \right\}\,.
\end{equation}
The equilibrium distance $r_0$ and spring constant $K$ take the same values as above, the maximum extension we set to $R_\mathrm{m} \equiv 0.1$.
Additionally, we introduce a hard-sphere potential between all pairs of nodes to ensure self-avoidance,
\begin{equation}
  V_\mathrm{hs} (r) =
  \begin{cases}
    0 & \textrm{if } r > 2 r_\mathrm{hs}\,, \\
    \infty & \textrm{if } r \leq 2 r_\mathrm{hs}\,, \\
  \end{cases}
\end{equation}
with hard-sphere radius $r_\mathrm{hs} \equiv 0.15$.
The configuration of the membrane is then described by the set of position vectors $\{\mathbf{r}_{k,l}\}_{(k,l) = (1,1), \dots, (L_x,L_y)}$.  The contribution of the membrane to the total energy of the system thus reads
\begin{eqnarray}
  \label{eq:mem_energy}
  E^\mathrm{mem} & = & \sum_{k=1}^{L_x-1} \sum_{l=1}^{L_y} V_\mathrm{FENE}^\mathrm{m} ( \left| \mathbf{r}_{k,l} - \mathbf{r}_{k+1,l} \right| ) \nonumber\\
  & & + \sum_{k=1}^{L_x} \sum_{l=1}^{L_y-1} V_\mathrm{FENE}^\mathrm{m} ( \left| \mathbf{r}_{k,l} - \mathbf{r}_{k,l+1} \right| )\,.
\end{eqnarray}
In our simulations, we set $L_x = L_y = 27$.  The polymer, which is anchored at the membrane center, can then take on all possible conformations and its fluctuations are not limited by the membrane boundaries.

The interaction between polymer and membrane is modeled by another LJ potential between all pairs of monomers and membrane nodes,
\begin{equation}
  \label{eq:V_LJ_pm}
  V_\mathrm{LJ}^\mathrm{pm} (r) =
  4 \epsilon_\mathrm{pm} \left[ \left( \sigma / r \right)^{12} - \left( \sigma / r \right)^6 \right] \,.
\end{equation}
The interaction strength between polymer and membrane is determined by the energy parameter $\epsilon_\mathrm{pm}$ which serves as an external control parameter, with values between $0.05$ and $1.50$.
Furthermore, between the first monomer and the central node acts the FENE potential $V_\mathrm{FENE}^\mathrm{p} (r)$, see Eq.~\eqref{eq:V_FENE_p}, as the anchoring potential.
The interaction term is the third contribution to the energy of the system,
\begin{eqnarray}
  \label{eq:int_energy}
  E^\mathrm{int} & = &
  \sum_{i=1}^{N} \sum_{k=1}^{L_x} \sum_{l=1}^{L_y} V_\mathrm{LJ}^\mathrm{pm} ( \left| \mathbf{r}_i - \mathbf{r}_{k,l} \right| ) \nonumber\\
  & & + V_\mathrm{FENE}^\mathrm{p} ( \left| \mathbf{r}_1 - \mathbf{r}_{L_x/2,L_y/2} \right| )\,,
\end{eqnarray}
yielding a total energy $E = E^\mathrm{pol} + E^\mathrm{mem} + E^\mathrm{int}$ entering the partition function $Z = \sum \exp (-\beta E)$ where $\beta \equiv 1/k_\mathrm{B}T$ is the inverse thermal energy and the summation extends over all possible microstates of the polymer and membrane.

For all the simulations presented in this work, we constrain the membrane to
fixed boundary conditions, that is, all nodes at the boundary of the membrane
are kept fixed.  Their positions are uniformly arranged on a rectangle of
dimensions $(L_x-1) r_0 \times (L_y-1) r_0$ in the $xy$ plane, such that the
membrane can adopt its ground state of a regular square lattice.
\revis{This is a simple choice which satisfies computational needs.  A different
  composition of the membrane mesh structure will doubtlessly have an influence on
  the location and specific properties of low-temperature crystalline pseudophases
  of the system.  Nonetheless, we expect that the qualitative structure of the
  pseudophase diagram, as obtained in our study, remains widely unchanged.}
The main feature of our membrane model is its flexibility which allows for thermal fluctuations around the planar ground state.  To identify the effects caused by this flexibility we also study the case where the whole membrane is kept fixed in its ground state.  We will distinguish these two situations by speaking of the ``flexible membrane'' and ``stiff membrane'' systems.

\subsection{\label{sec:method}Simulation method}
The parallel tempering (PT) Monte Carlo algorithm \cite{geyer_annealing_1995, hukushima_exchange_1996} provides a conceptually simple but efficient method for the simulation of complex systems over a broad temperature range.  The basic idea is to perform parallel simulations on a number of identical copies (replicas) of the system at different temperatures using a standard sampling scheme (e.g., the Metropolis algorithm \cite{metropolis_equation_1953}).  After a certain number of sweeps (Monte Carlo ``time'' steps) the conformations are exchanged with an acceptance probability
\begin{equation}
  A_\mathrm{PT} = \begin{cases}
    1 & \textrm{for } \Delta < 0 \\
    \exp(-\Delta) & \textrm{for } \Delta > 0 \,,
  \end{cases}
\end{equation}
where $\Delta = (\beta_n - \beta_m) [E(\mathbf{X}) - E(\mathbf{X}')]$.  $\mathbf{X}$ and $\mathbf{X}'$ are the conformations of the replicas at inverse thermal energy $\beta_m$ and $\beta_n$, respectively.  If the temperature values of the PT simulation are chosen thoroughly \cite{bittner_make_2008}, the conformations will perform a random walk in temperature space and therefore rapidly decorrelate, helping to escape from trapped states.

The density of states and canonical expectation values for temperatures in between the selected values can be obtained by the multiple-histogram reweighting technique \cite{ferrenberg_optimized_1989}.

\subsection{\label{sec:observables}Observables}
To identify and distinguish the different structural phases of the system we define a number of observable quantities $\mathcal{O}$.  The behavior of the canonical expectation values $\langle\mathcal{O}\rangle$ and their temperature derivatives $\partial \langle\mathcal{O}\rangle / \partial T = k_\mathrm{B} \beta^2 [\langle\mathcal{O}E\rangle - \langle\mathcal{O}\rangle \langle E \rangle]$ will provide information about the structural phases and transitions.

The most basic quantities are the canonical averages of the total energy $\langle E \rangle$ and its individual contributions $\langle E^\mathrm{pol} \rangle$, $\langle E^\mathrm{mem} \rangle$, and $\langle E^\mathrm{int} \rangle$ together with the associated heat capacities (temperature derivatives) $C$, $C^\mathrm{pol}$, $C^\mathrm{mem}$, and $C^\mathrm{int}$, where $C^\mathrm{pol} = \partial \langle E^\mathrm{pol} \rangle / \partial T$, and so on.  Evidently, $\langle E \rangle = \langle E^\mathrm{pol} \rangle + \langle E^\mathrm{mem} \rangle + \langle E^\mathrm{int} \rangle$ and $C = C^\mathrm{pol} + C^\mathrm{mem} + C^\mathrm{int}$.

The radius of gyration $R_\mathrm{g}$ is a measure for the overall compactness of the polymer conformation.  It is defined as the root-mean-square distance of the individual monomers from the center of mass of the polymer $\mathbf{r}_\mathrm{c.m.} = N^{-1} \sum_{n=1}^N \mathbf{r}_n$, $\langle R_\mathrm{g} \rangle = \langle [N^{-1} \sum_{n=1}^{N} ( \mathbf{r}_n - \mathbf{r}_\mathrm{c.m.} )^2 ]^{1/2} \rangle$.  Although the membrane surface is not planar, it may be instructive to separate the radius of gyration into the components of the gyration tensor parallel and perpendicular to the membrane equilibrium state, i.e., to the $xy$-plane, $\langle R_\mathrm{g,\parallel} \rangle = \langle \{ N^{-1} \sum_{n=1}^{N} [ ( x_n - x_\mathrm{c.m.} )^2 + ( y_n - y_\mathrm{c.m.} )^2 ] \}^{1/2} \rangle$ and $\langle R_\mathrm{g,\perp} \rangle = \langle [ N^{-1} \sum_{n=1}^{N} ( z_n - z_\mathrm{c.m.} )^2 ]^{1/2} \rangle$.  The ratio of these two values gives the sphericity aspect ratio with respect to the $xy$-plane $\Psi_\mathrm{r} = \sqrt{2} \langle R_\mathrm{g,\perp} \rangle / \langle R_\mathrm{g,\parallel} \rangle$.  This definition is chosen such that $\Psi_\mathrm{r} = 1$ indicates spherically symmetric structures, while an oblate (prolate) spheroid with polar axis $z$ will have $\Psi_\mathrm{r} < 1$ ($\Psi_\mathrm{r} > 1$).

To determine whether the polymer is, on average, close to the membrane surface (adsorbed) or freely exploring the third dimension (desorbed), we measure the distance of the center of mass of the polymer from the membrane equilibrium state in the $xy$ plane (i.e., its $z$ component, $\langle z_\mathrm{c.m.} \rangle = \langle N^{-1} \sum_{n=1}^{N} z_n \rangle$).

The number of contacts between monomers and membrane nodes may give excellent information about the state of adsorption.  As a reasonable, but still to some extent arbitrary, measure for a ``contact'' we decided to count every monomer-node pair that contributes a $V_\mathrm{LJ}^\mathrm{pm}(r)$ energy less than a threshold value $E_\mathrm{c}^\mathrm{pm} \equiv -0.5 \epsilon_\mathrm{pm}$ as a contact yielding a mean number of polymer-membrane contacts $\langle n_\mathrm{pm} \rangle = \langle N^{-1} \sum_{i=1}^{N} \sum_{k=1}^{L_x} \sum_{l=1}^{L_y} \Theta ( E_\mathrm{c}^\mathrm{pm} - V_\mathrm{LJ}^\mathrm{pm} ( \left| \mathbf{r}_i - \mathbf{r}_{k,l} \right| ) ) \rangle$, where $\Theta(x)$ denotes the Heaviside step function.  Similarly, we define the number of intrinsic contacts as another measure for the compactness of the polymer, $\langle n_\mathrm{pp} \rangle = \langle N^{-1} \sum_{i=1}^{N-1} \sum_{j=i+1}^{N} \Theta [ E_\mathrm{c}^\mathrm{pp} - V_\mathrm{LJ}^\mathrm{pp} ( \left| \mathbf{r}_i - \mathbf{r}_{j} \right| ) ] \rangle$, with $E_\mathrm{c}^\mathrm{pp} \equiv -0.5 \epsilon_\mathrm{pp}$.

\section{\label{sec:results}Results and discussion}
\subsection{\label{sec:stiff}Stiff membrane system}
The stiff membrane system is studied as a reference system here.  On the one hand, the knowledge about this system will enable us to point out the differences in the behavior of the stiff and the flexible membrane cases and to identify the new effects emerging from the surface flexibility.  On the other hand, the stiff membrane case allows a comparison of our model with studies on polymer adsorption to solid (flat) substrates \cite{moeddel_conformational_2009, moeddel_systematic_2010}.
For the stiff membrane system we keep all membrane nodes, in agreement with the boundary conditions, fixed on a regular square lattice with lattice spacing $r_0$ in the $xy$ plane.  This is obviously the ground state of the membrane as all FENE springs are at equilibrium, $V_\mathrm{FENE}^\mathrm{m} (r_0) = 0 $, $E^\mathrm{mem} = 0$, and any deviation from this state would introduce a positive contribution to the membrane energy.

The main results we show here were obtained by parallel tempering simulations with 24 replicas in the temperature range from $0.021$ to $1.500$ and $8 \times 10^6$ sweeps on each replica.  Exchanges of conformations between the replica were attempted every $20$ sweeps.  For the uncritical region above $T=1.5$, additional PT simulations with $10^6$ sweeps were carried out.  All the simulations were performed at $30$ different values of $\epsilon_\mathrm{pm}$ in the interval $\epsilon_\mathrm{pm} = 0.05, \dots, 1.50$.

\subsubsection{Pseudophase diagram}
\begin{figure}
  \includegraphics[width=\linewidth]{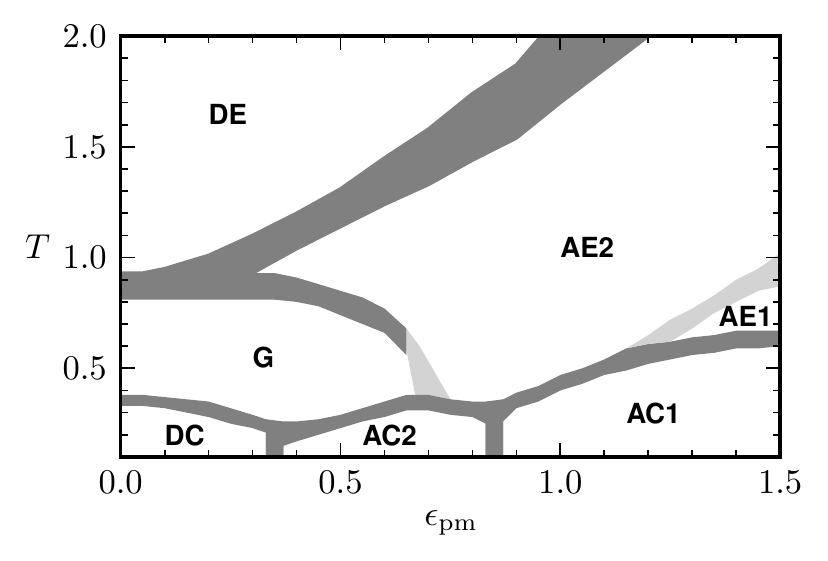}
  \caption[Pseudophase diagram for the stiff membrane system]{Pseudophase diagram for the stiff membrane system.  The main structural phases divide into predominantly adsorbed (\textsf{A}), and desorbed (\textsf{D}) on the one hand and expanded (\textsf{E}), globular (\textsf{G}), and compact (\textsf{C}) structures on the other hand.}
  \label{fig:ppd_mf0}
\end{figure}
The main information about the structural behavior is summarized in the pseudo\-phase diagram shown in Fig.~\ref{fig:ppd_mf0}.  It displays the structural pseudophases and pseudo-phase transitions in the $\epsilon_\mathrm{pm}$-$T$-plane.  Temperature increases from bottom to top, interaction strength increases from left to right.

\begin{figure}
  \centering
  \includegraphics[width=\linewidth]{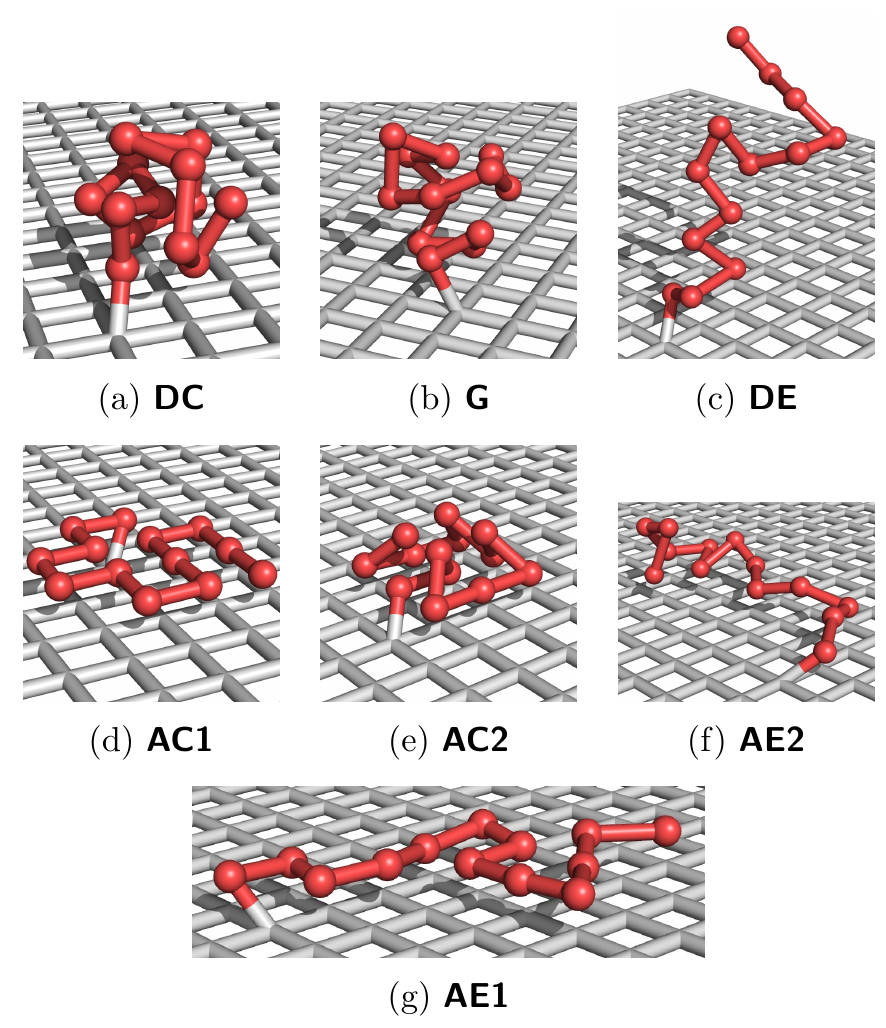}
  \caption[Phases of the stiff membrane system]
  {(Color online) Representative conformations in the individual phases of the stiff membrane system (see text for a detailed description).}
  \label{fig:conf_mf0}
\end{figure}

The structural phases in which certain conformations predominate are labeled by a letter code adopted from Ref.~\cite{moeddel_conformational_2009}.  Since we are working with a finite system, we cannot identify precise transition lines but rather transition regions between the phases.  These are displayed as the shaded regions in the phase diagram where the dark shades indicate well-founded transitions and the bright shades stand for less established transitions.  In the following, we describe all the identified structural pseudophases, for pictures of typical conformations in each phase, see Fig.~\ref{fig:conf_mf0}.
\begin{description}
\item[Desorbed Compact \textsf{(DC)}] The most compact conformations the polymer can adopt.  For the example considered here ($N = 13$), these are the highly ordered icosahedral states we know from the free polymer \cite{schnabel_elastic_2009}.  Here, it is simply attached to the membrane surface [Fig.~\ref{fig:conf_mf0}(a)].
\item[Globular \textsf{(G)}] Compact but disordered conformations.  Similar to \textsf{DC} these are globular states of the free polymer attached to the membrane.  Because of the short chain length, we cannot distinguish here between adsorbed and desorbed globular conformations [Fig.~\ref{fig:conf_mf0}(b)].
\item[Desorbed Expanded \textsf{(DE)}] The most disordered conformations.  These are random coil structures of the free polymer anchored to the membrane and restricted to the upper half-space [Fig.~\ref{fig:conf_mf0}(c)].
\item[Adsorbed Compact, Single Layer \textsf{(AC1)}] Completely adsorbed, compact conformations.  These film-like structures are constrained to lie flat on the surface of the membrane and to its square lattice structure.  Additionally they are mainly disk-like compact [Fig.~\ref{fig:conf_mf0}(d)].
\item[Adsorbed Compact, Double Layer \textsf{(AC2)}] Partially adsorbed, compact conformations.  These consist of a bottom layer adsorbed to the membrane and adapted to its lattice structure and a second layer on top of it building a semi-spherical droplet [Fig.~\ref{fig:conf_mf0}(e)].
\item[Adsorbed Expanded, Double Layer \textsf{(AE2)}] Partially adsorbed, extended conformations.  These are random coil structures mainly adsorbed to the membrane surface but also extending into the third dimension [Fig.~\ref{fig:conf_mf0}(f)].
\item[Adsorbed Expanded, Single Layer \textsf{(AE1)}] Completely adsorbed, extended conformations.  These are mainly two-dimensional random coil structures adsorbed to the membrane surface [Fig.~\ref{fig:conf_mf0}(g)].
\end{description}

\subsubsection{Structural phases and transitions}

In the following, we substantiate the proposed pseudophases and discuss the transitions by looking in detail at the observables measured in the course of the simulations.

The simulation methods we apply here are not specialized in finding ground states.  For this task simpler and more efficient algorithms exist.  Nevertheless, our simulation results reach down to sufficiently low temperatures where a convergence sets in.  We therefore think that the low-energy conformations we found in the low-temperature phases \textsf{DC}, \textsf{AC2}, and \textsf{AC1} are good approximations of the real ground states.  Of course, further details like the actual bond distribution, translations on the membrane surface, and distortions of the idealized symmetries due to LJ interactions over longer ranges have to be considered for more precise ground-state predictions.

The different compact conformations are characterized in a simple way by the contact numbers (see Sec.~\ref{sec:observables}).  The icosahedron is the most compact structure of the $13$mer in \textsf{DC}, resulting in the highest number of intrinsic contacts.  Counting them correctly we end up with $n_\mathrm{pp}^\textsf{DC} = 42/13 \approx 3.23 $.  Due to its compactness, this structure can build up few polymer-membrane contacts, $n_\mathrm{pm}^\textsf{DC} \approx 1$.

\begin{figure}
  \centering
  \includegraphics[width=\linewidth]{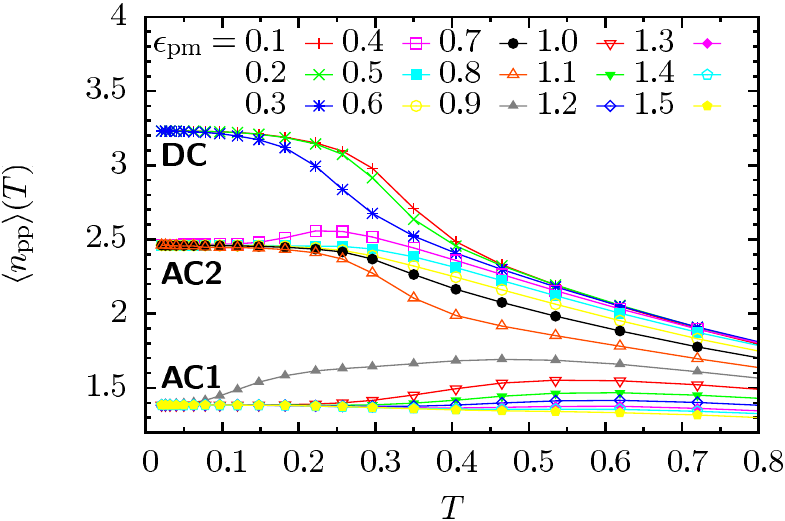}\\
  \includegraphics[width=\linewidth]{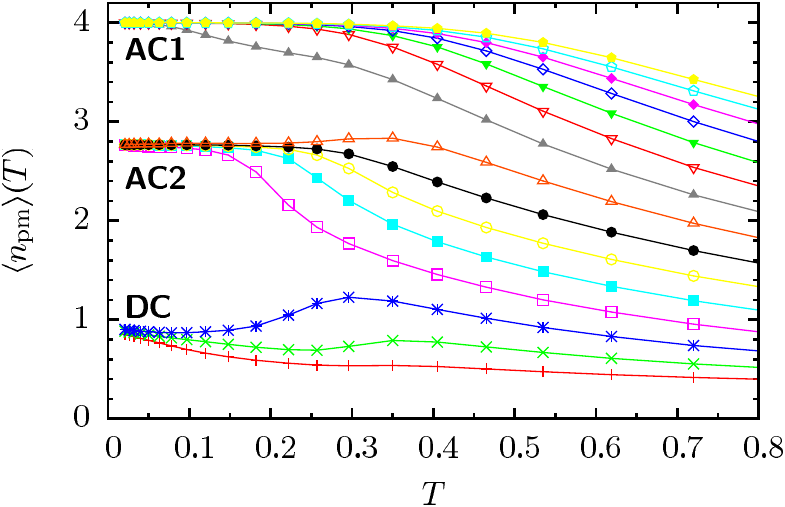}
  \caption[Mean contact numbers in the stiff membrane system]{(Color online) Mean number of intrinsic (top) and polymer-membrane (bottom) contacts as a function of temperature for various polymer-membrane interaction strengths in the stiff membrane system.  The legend applies to both plots.}
  \label{plot:mf0_lowT_n}
\end{figure}

In the two-dimensional compact disk-like structures of \textsf{AC1} every monomer is in contact with four membrane nodes, $n_\mathrm{pm}^\textsf{AC1} = 4$, and we count $n_\mathrm{pp}^\textsf{AC1} = 18/13 \approx 1.38$ intrinsic contacts per monomer.  In the partially adsorbed semi-spherical droplets of \textsf{AC2}, nine of the monomers are in contact with four membrane nodes each, $n_\mathrm{pm}^\textsf{AC2} = 36/13 \approx 2.78$, for the intrinsic contacts we get $n_\mathrm{pp}^\textsf{AC2} = 32/13 \approx 2.46$.  Equipped with this information, we can easily identify the different low-temperature phases.  Figure~\ref{plot:mf0_lowT_n} shows the mean numbers of intrinsic and polymer-membrane contacts for different polymer-membrane interaction strengths over the whole interaction strength spectrum investigated here.  At low temperatures, we see convergence to exactly the values predicted above.

For $\epsilon_\mathrm{pm} \leq 0.3$, the mean numbers of intrinsic contacts approach the value $3.23$ uniquely characterizing \textsf{DC}.  The mean number of polymer-membrane contacts does not exhibit such a clear convergence in this case, since reorientation on top of the membrane is still possible in this phase.
For $0.4 \leq \epsilon_\mathrm{pm} \leq 0.8$, the mean contact numbers approach $2.46$ for the intrinsic contacts and $2.78$ for the polymer-membrane contacts.  Together, these values identify compact double-layer conformations as approximate ground states in this phase (\textsf{AC2}).
For $\epsilon_\mathrm{pm} \geq 0.9$, the mean numbers of $1.38$ intrinsic contacts and four polymer-membrane contacts are approached for low temperatures.  We therefore find monolayer structures as approximate ground states in this region (\textsf{AC1}).

For low polymer-membrane interaction strength, the polymer should behave similarly to the free polymer \cite{schnabel_surface_2009,schnabel_elastic_2009}.  The freezing transition between the compact icosahedral and the globular phase is identified by peaks in the polymer heat capacity and temperature derivative of the radius of gyration around $T = 0.35$.  The latter quantity shows another peak at around $T = 0.9$ indicating the $\Theta$ transition between the globular and expanded phase.
\begin{figure}
  \centering
  \includegraphics[width=\linewidth]{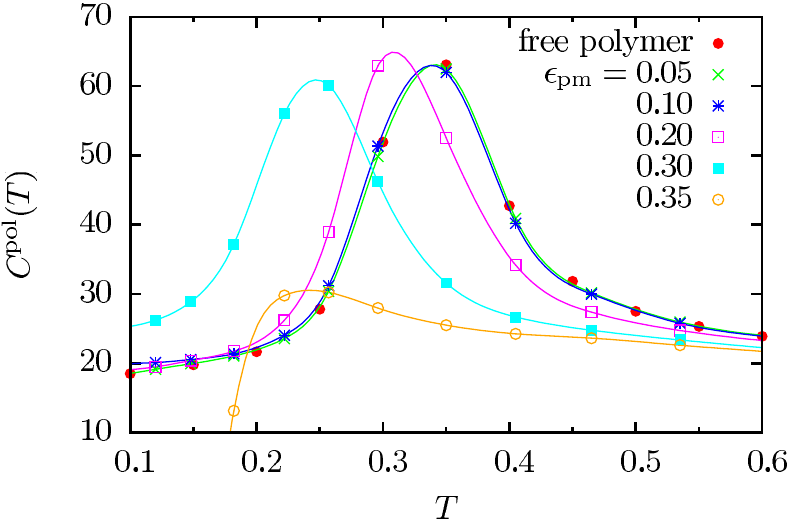}\\
  \includegraphics[width=\linewidth]{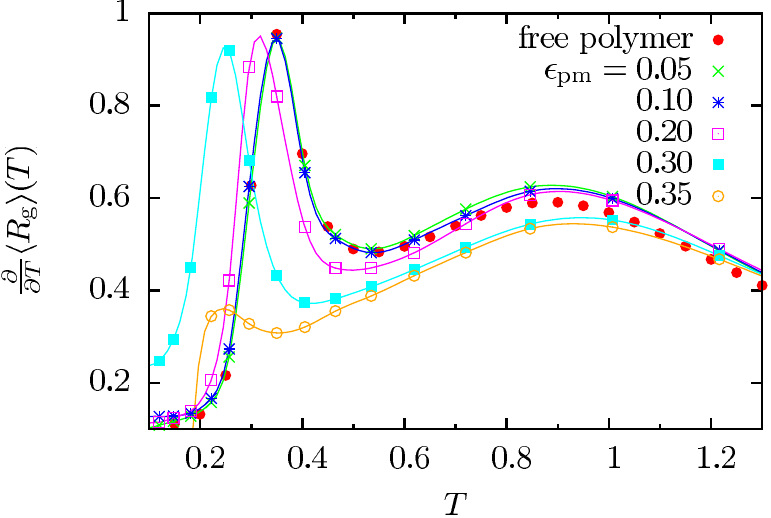}
  \caption[Polymer heat capacity and temperature derivative of the radius of gyration for low polymer-membrane interaction strengths in the stiff membrane system]
  {(Color online) Polymer heat capacity (top) and temperature derivative of the radius of gyration (bottom) for low polymer-membrane interaction strengths in the stiff membrane system.  For better comparison, the values for the free polymer are also plotted (red circles).}
  \label{plot:mf0_dc-g-de}
\end{figure}
In Fig.~\ref{plot:mf0_dc-g-de}, we plot these two quantities for the stiff membrane system.  The curves for $\epsilon_\mathrm{pm} = 0.05$ and $0.10$ lie almost on top of each other and agree almost perfectly with those for the free polymer (red circles).  We conclude that the same kinds of transitions take place here, from \textsf{DC} over \textsf{G} to \textsf{DE}.  For $\epsilon_\mathrm{pm} = 0.20$ and $0.30$ the freezing peaks shift to lower temperatures in both quantities, disappearing almost completely at $\epsilon_\mathrm{pm} = 0.35$ which marks the point where the ground state changes from \textsf{DC} to \textsf{AC2} (see Fig.~\ref{plot:mf0_lowT_n}).  The $\Theta$ peak maintains its position $T \approx 0.9$ in the whole range of $0.05 \leq \epsilon_\mathrm{pm} \leq 0.35$.

\begin{figure}
  \centering
  \includegraphics[width=\linewidth]{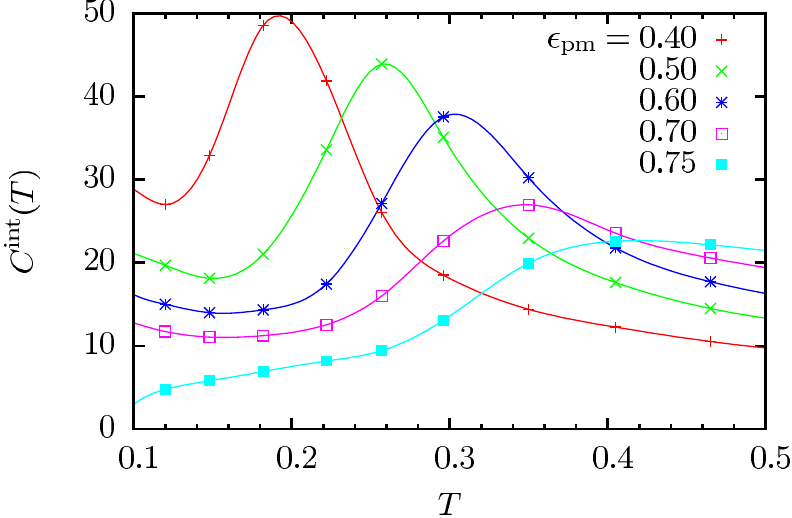}\\
  \includegraphics[width=\linewidth]{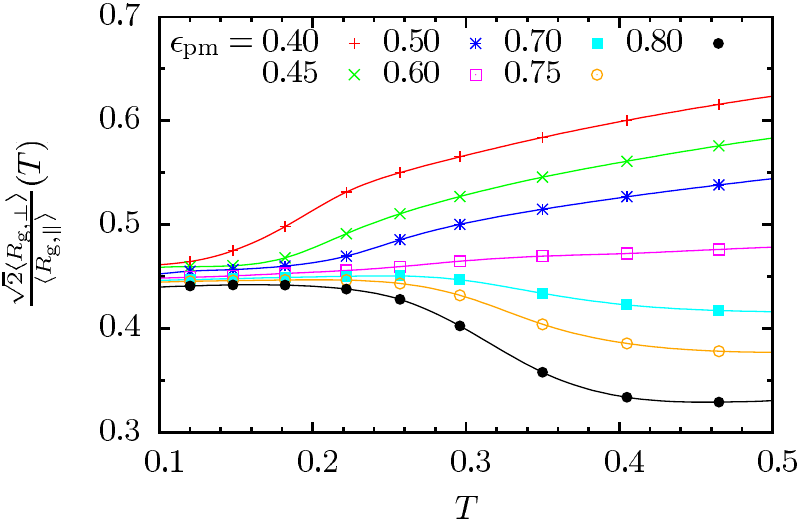}
  \caption[Contribution of polymer-membrane interaction to heat capacity and sphericity aspect ratio at the \textsf{AC2}~--~\textsf{G} transition in the stiff membrane system]
  {(Color online) Contribution of polymer-membrane interaction to heat capacity (top) and sphericity aspect ratio (bottom) at the \textsf{AC2}~--~\textsf{G} transition in the stiff membrane system.}
  \label{plot:mf0_ac2-g}
\end{figure}
In the range of $0.4 \lesssim \epsilon_\mathrm{pm} \lesssim 0.6$, the mean polymer-membrane contacts decrease considerably between $T \approx 0.15$ and $0.35$ without a substantial change in the intrinsic contacts (Fig.~\ref{plot:mf0_lowT_n}) suggesting a (partial) desorption transition into the globular phase \textsf{G}.  This assumption is supported by a peak in the interaction part of the heat capacity $C^\mathrm{int}(T)$, indicating that the interaction energy becomes higher in this region, and an increase of the sphericity aspect ratio in that range (Fig.~\ref{plot:mf0_ac2-g}).  We find that the transition temperature becomes higher for increasing interaction strength, from $T \approx 0.18$ at $\epsilon_\mathrm{pm} = 0.4$ to $T \approx 0.30$ at $\epsilon_\mathrm{pm} = 0.6$.  For larger values of the interaction strength, we see the sphericity aspect ratio decreasing which suggests a transition into structures extending parallel to the membrane.

\begin{figure}
  \centering
  \includegraphics[width=\linewidth]{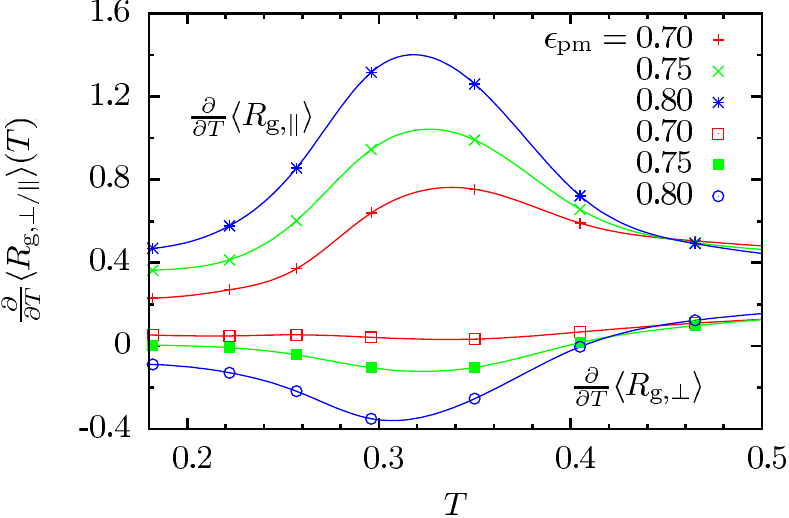}
  \caption[Temperature derivatives of the components of the gyration tensor at the \textsf{AC2}~--~\textsf{AE2} transition in the stiff membrane system]
  {(Color online) Temperature derivatives of the parallel (upper curves) and perpendicular (lower curves) component of the gyration tensor at the \textsf{AC2}~--~\textsf{AE2} transition in the stiff membrane system.}
  \label{plot:mf0_ac2-ae2_dtrg}
\end{figure}
This can be affirmed by looking at the two components of the gyration tensor separately.  Figure~\ref{plot:mf0_ac2-ae2_dtrg} displays the temperature derivatives of the two components for $0.7 \leq \epsilon_\mathrm{pm} \leq 0.8$.  We see a clear peak in the parallel component (upper three curves) and no signal or a valley in the perpendicular component (lower three curves) around $0.30 \lesssim T \lesssim 0.35$ meaning that the polymer does extend in the $xy$ direction while it does not, or even contracts, in the $z$ direction.  This strongly supports the assumption that we have a transition from the droplet phase \textsf{AC2} to the adsorbed three-dimensional random coil phase \textsf{AE2} here.

\begin{figure}
  \centering
  \includegraphics[width=\linewidth]{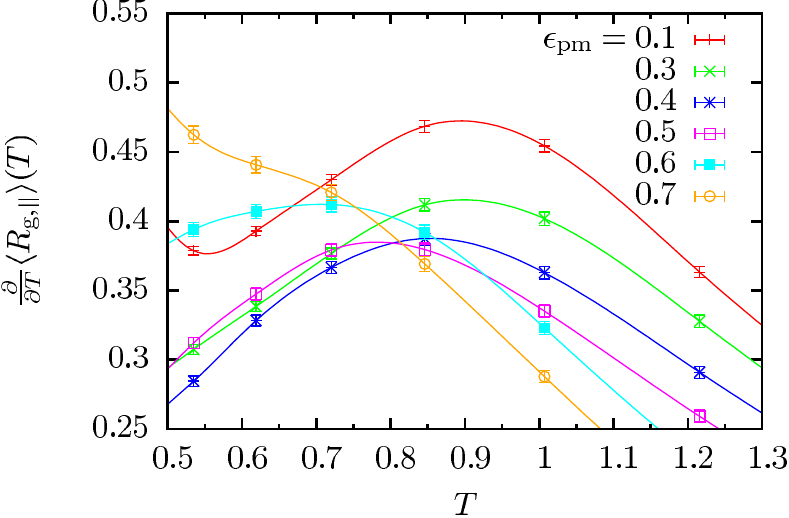}
  \caption{(Color online) Temperature derivative of the parallel component of the gyration tensor at the \textsf{G}~--~\textsf{AE2} transition in the stiff membrane system.}
  \label{plot:mf0_g-ae2_dtrgxy}
\end{figure}
The transition from the globular phase \textsf{G} to the adsorbed expanded, double layer, phase \textsf{AE2} is difficult to identify since globular structures are due to the anchoring approximately as close to the surface as the adsorbed but still three-dimensional random coils of \textsf{AE2}.  Nevertheless, we do observe an increase in the parallel component of the gyration tensor indicated by a peak in its temperature derivative for $\epsilon_\mathrm{pm} \lesssim 0.6$ (Fig.~\ref{plot:mf0_g-ae2_dtrgxy}).  For increasing $\epsilon_\mathrm{pm}$, the peak position slightly moves to lower temperatures and disappears from $\epsilon_\mathrm{pm} \approx 0.7$ on.  We will see below that the perpendicular component experiences a similar increase, which for $0.3 \lesssim \epsilon_\mathrm{pm} \lesssim 0.6$ takes place at significantly higher temperatures.  Therefore, we conclude that the \textsf{G}~--~\textsf{DE} transition discussed above separates for higher interaction strengths into two transitions, an expansion \textsf{G}~--~\textsf{AE2} and a desorption \textsf{AE2}~--~\textsf{DE}.

\begin{figure}
  \centering
  \includegraphics[width=\linewidth]{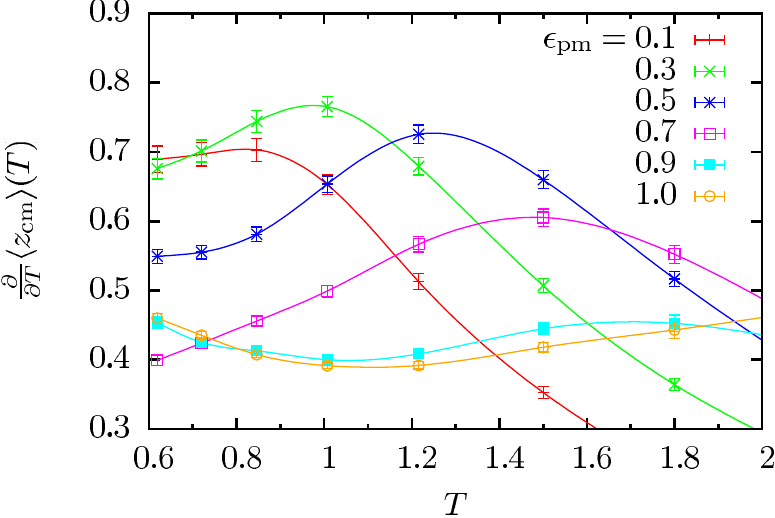}\\
  \includegraphics[width=\linewidth]{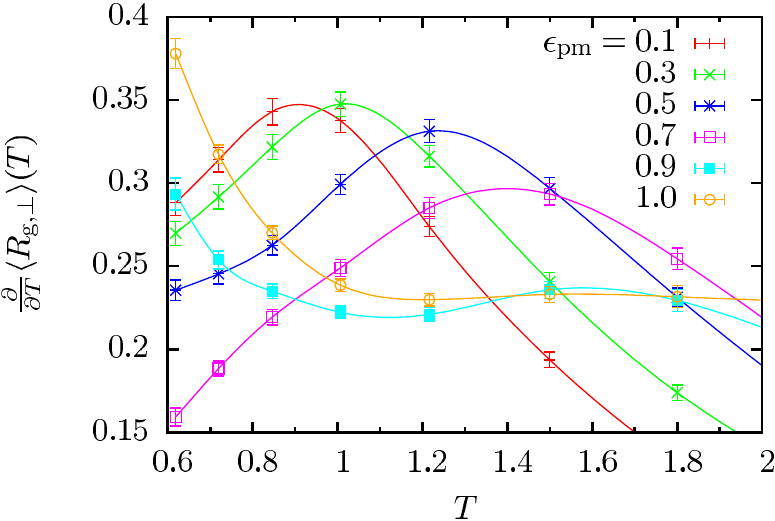}
  \caption[Temperature derivative of the $z$ component of the center of mass and perpendicular component of the gyration tensor at the \textsf{AE2}~--~\textsf{DE} transition in the stiff membrane system]
  {(Color online) Temperature derivative of the $z$ component of the center of mass (top) and perpendicular component of the gyration tensor (bottom) at the \textsf{AE2}~--~\textsf{DE} transition in the stiff membrane system.}
  \label{plot:mf0_ae2-de}
\end{figure}
The suggested desorption transition from adsorbed three-dimensional random coils \textsf{AE2} to desorbed random coils \textsf{DE} should be measurable by a simultaneous increase in the perpendicular component of the gyration tensor and the height of the polymer's center of mass above the membrane surface.  These two effects can indeed be observed as peaks in the respective temperature derivatives (Fig.~\ref{plot:mf0_ae2-de}).  As in many transitions of finite systems, the peak positions slightly differ from each other causing the rather broad transition region in the phase diagram.

\begin{figure}
  \centering
  \includegraphics[width=\linewidth]{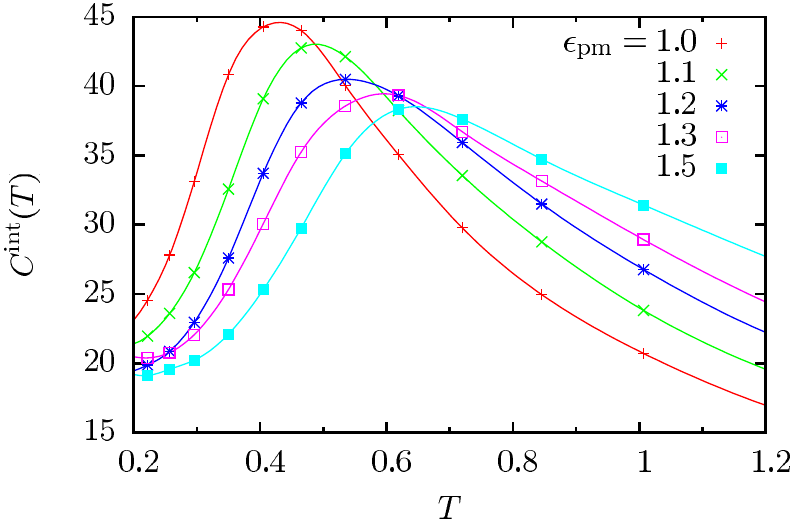}
  \caption[Interaction heat capacity at the \textsf{AC1}~--~\textsf{AE2} transition in the stiff membrane system]
  {(Color online) Interaction heat capacity at the \textsf{AC1}~--~\textsf{AE2} transition in the stiff membrane system.}
  \label{plot:mf0_ac1-ae_cint}
\end{figure}

\begin{figure}
  \centering
  \includegraphics[width=\linewidth]{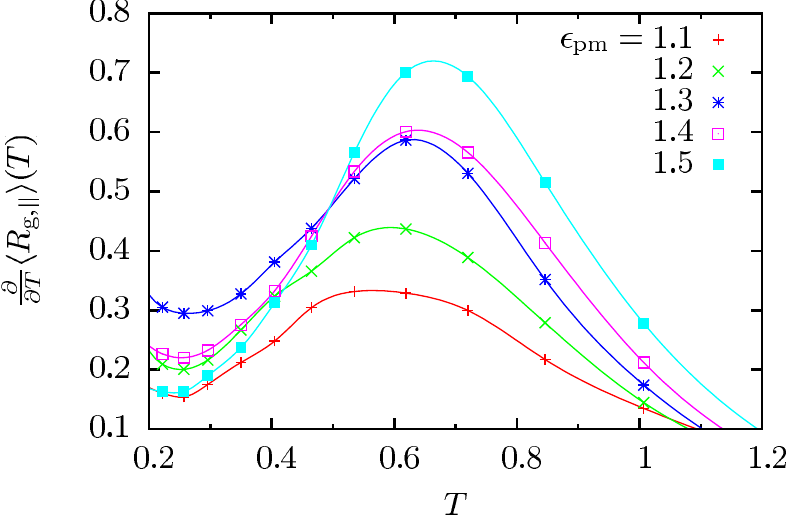}\\
  \includegraphics[width=\linewidth]{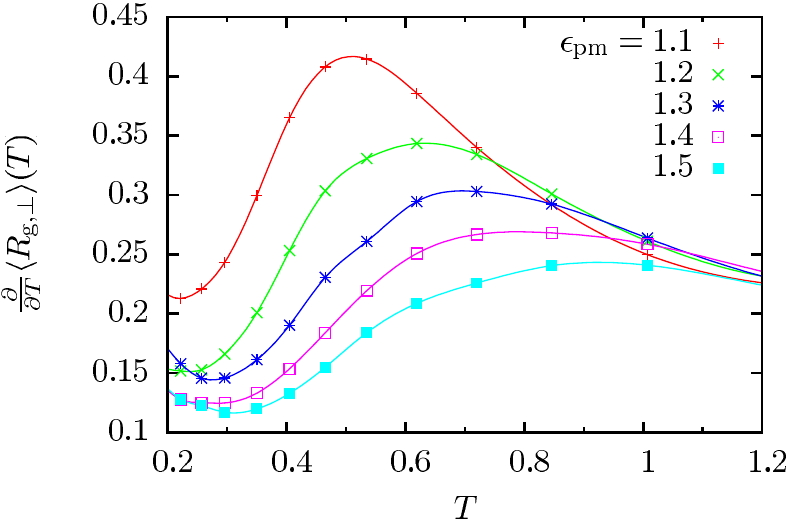}
  \caption[Temperature derivative of the components of the gyration tensor at the \textsf{AC1}~--~\textsf{AE2} / \textsf{AE1} transition in the stiff membrane system]
  {(Color online) Temperature derivative of the parallel (top) and perpendicular (bottom) components of the gyration tensor at the \textsf{AC1}~--~\textsf{AE2} / \textsf{AE1} transition in the stiff membrane system.}
  \label{plot:mf0_ac1-ae}
\end{figure}
We conclude this section by discussing the last transition between the compact film phase \textsf{AC1} and the adsorbed expanded phases \textsf{AE2} and \textsf{AE1}.  Figure~\ref{plot:mf0_ac1-ae_cint} shows the interaction heat capacity for high polymer-membrane interaction strengths $1.0 \leq \epsilon_\mathrm{pm} \leq 1.5$.  We see a clear peak indicating a (partial) desorption transition shifting to higher temperatures for stronger polymer-membrane interaction.  This is supported by peaks in the temperature derivatives of the components of the gyration tensor (Fig.~\ref{plot:mf0_ac1-ae}).  Taking a closer look at these figures, we see that with increasing interaction strength the parallel expansion shifts more slowly to higher temperatures than the expansion perpendicular to the membrane.  Although this disagreement might be another finite-size effect, we think that it is caused by the onset of the completely adsorbed random coil phase \textsf{AE1}.  The existence of this phase is also supported by the fact that the parallel expansion becomes stronger while the perpendicular expansion becomes weaker for higher polymer interaction strength, as suggested by the peak heights.

\subsubsection{Comparison with established results}

The phase diagram we have constructed here (Fig.~\ref{fig:ppd_mf0}) qualitatively compares with that by M\"oddel \textit{et al.} \cite{moeddel_conformational_2009}, Fig. 2, which allowed us to adopt their notation.  The main qualitative differences between the models introduced there and here are the anchoring of the polymer and the discrete surface structure.  For this reason, we do not expect quantitative agreement in the locations of the transitions but qualitatively, both systems should behave similarly.

As mentioned above we cannot distinguish between adsorbed and desorbed globular structures in our case, so we just have a single globular phase \textsf{G} in contrast to the adsorbed globular \textsf{AG} and desorbed globular \textsf{DG} phases in Ref.~\cite{moeddel_conformational_2009}.  Also we do not find an adsorbed compact phase, like \textsf{AC2a} in Ref.~\cite{moeddel_conformational_2009}, that can be clearly distinguished from \textsf{DC} and \textsf{AC2}.  Obviously these two differences stem from the anchoring of the polymer.  For longer chains the distinctions possibly can be recovered also in the grafted case.

Another difference can be identified in the transition line between compact and globular or expanded structures.  In Ref.~\cite{moeddel_conformational_2009}, these transitions occured at more or less constant temperature, independently of the polymer-surface interaction strength.  This seems to be not the case in our model.

\subsection{Flexible membrane system}

We now extend the focus of this study and discuss the structural behavior of the coupled system consisting of an elastic polymer anchored to a flexible membrane.  This section describes our findings about the flexible membrane system in detail where we concentrate on the differences and similarities with the stiff membrane case.

All results presented here were obtained by parallel tempering simulations with 48 replicas in the temperature range from $0.055$ to $1.500$ and $10^7$ sweeps on each replica.  Exchanges of conformations between the replica were attempted every $20$ sweeps.  The simulations were carried out at $30$ different values of $\epsilon_\mathrm{pm} = 0.05, \dots, 1.50$.  As the performance of the parallel tempering conformation exchanges turned out to be poor in the temperature range below $T \approx 0.1$ leading to large statistical errors, we restrict our discussion to temperatures above this value.

\subsubsection{Pseudophase diagram}

\begin{figure}
  \centering
  \includegraphics[width=\linewidth]{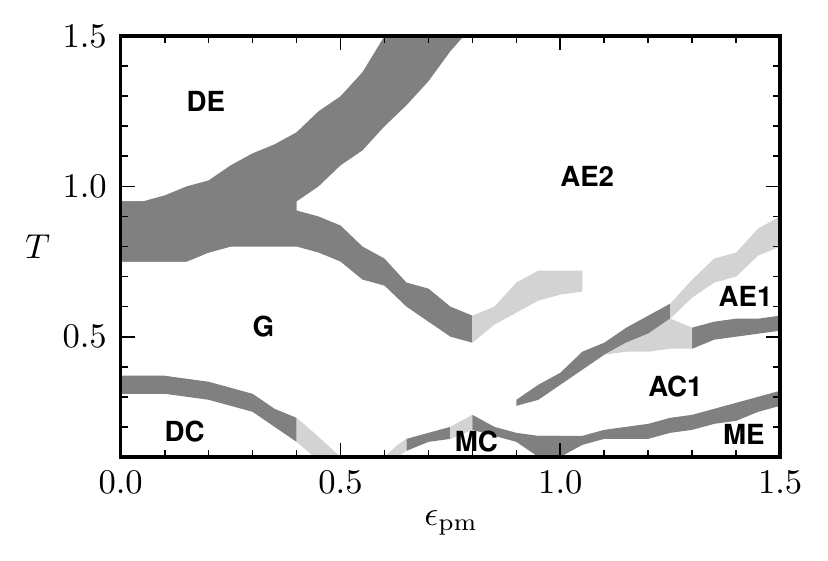}
  \caption[Pseudophase diagram for the flexible membrane system]
  {Pseudophase diagram for the flexible membrane system.  The main structural phases divide into predominantly adsorbed (\textsf{A}), desorbed (\textsf{D}), and embedded (\textsf{M}) on the one hand and expanded or elongated (\textsf{E}), globular (\textsf{G}), and compact (\textsf{C}) structures on the other hand.}
  \label{fig:ppd_mf1}
\end{figure}

The pseudophase diagram shown in Fig.~\ref{fig:ppd_mf1} summarizes the main information about the structural behavior of the flexible membrane system.  It displays the structural pseudophases and pseudo-phase transitions in the $\epsilon_\mathrm{pm}$-$T$ plane.  As before, temperature increases from bottom to top and polymer-membrane interaction strength from left to right.

\begin{figure}
  \centering
  \includegraphics[width=\linewidth]{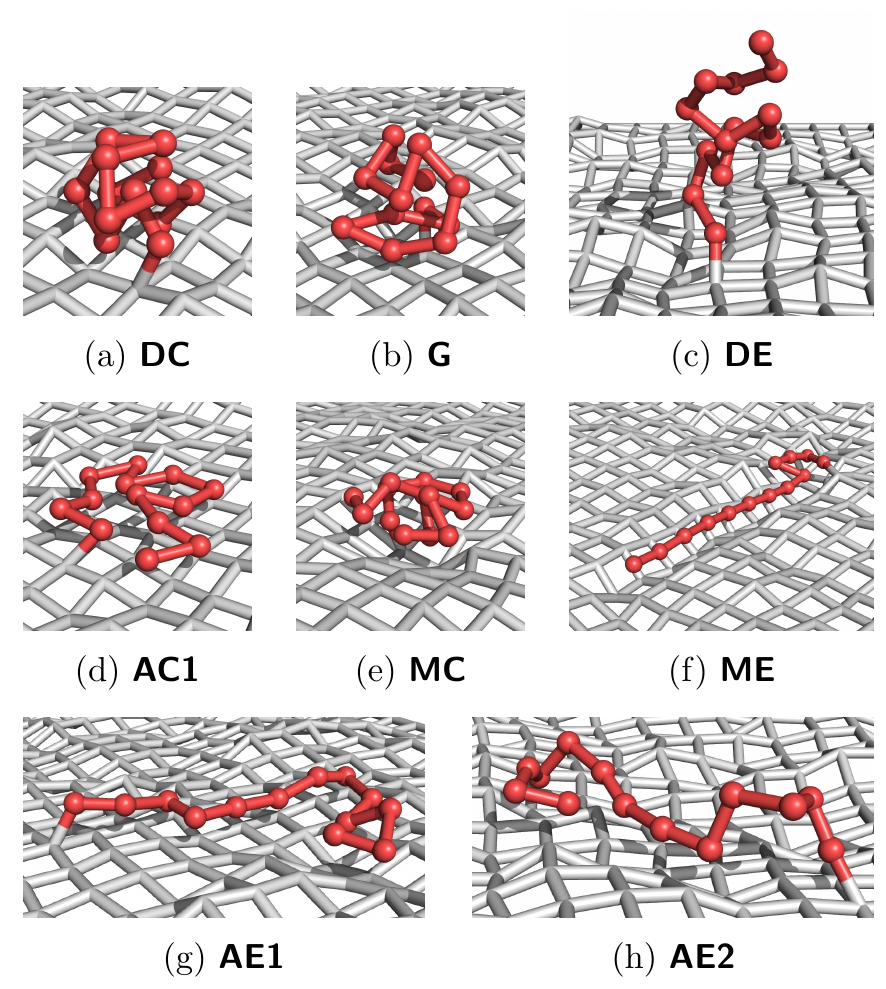}
  \caption[Phases of the flexible membrane system]
  {(Color online) Representative conformations in the individual phases of the flexible membrane system (see text for a detailed description).}
  \label{fig:conf_mf1}
\end{figure}

Again, the structural phases are labeled by a letter code similar to the one used before and transition regions are shaded in dark gray for well-founded and light gray for less certain transitions.  In the following, we describe the identified pseudophases in the flexible membrane system.  Images of typical conformations in each phase are shown in Fig.~\ref{fig:conf_mf1}.  The phases in which the polymer-membrane interaction is not of particular importance (\textsf{DC}, \textsf{G}, and \textsf{DE} [Fig.~\ref{fig:conf_mf1}(a)-(c)]) are, of course, very similar to the case of the stiff membrane.  To avoid redundancy we omit a description of these here.  The main new structures are the embedded conformations (\textsf{MC}, \textsf{ME}) which reflect the back-reaction between polymer and membrane fluctuations.
Let us now first summarize the differences in the phase behavior, compared to the stiff membrane case.
\begin{description}
\item[Adsorbed Compact, Single Layer \textsf{(AC1)}] Completely adsorbed, compact conformations.  These film-like structures lie flat on the membrane surface.  Since the membrane is deformable in this case, the conformations are, strictly speaking, not two-dimensional as before and do not necessarily adopt any lattice structure [Fig.~\ref{fig:conf_mf1}(d)].
\item[Embedded Compact \textsf{(MC)}] Compact conformations, deforming the membrane.  These highly compact structures are partially wrapped in by the membrane which, on the other hand, forms an immersion to incorporate the collapsed polymer [Fig.~\ref{fig:conf_mf1}(e)].
\item[Embedded Elongated \textsf{(ME)}] Stretched conformations, incorporated by the membrane.  For these structures the membrane forms a channel into which the almost fully stretched polymer is embedded.  The polymer is not randomly expanded here, but specifically elongated [Fig.~\ref{fig:conf_mf1}(f)].
\item[Adsorbed Expanded, Single Layer \textsf{(AE1)}] Completely adsorbed, extended conformations.  These are randomly expanded structures that live on the membrane surface.  In contrast to the stiff membrane case, the structures are not strictly two-dimensional as the membrane is fluctuating in all directions [Fig.~\ref{fig:conf_mf1}(g)].
\item[Adsorbed Expanded, Double Layer \textsf{(AE2)}] Partially adsorbed, extended conformations.  These are partly adsorbed random coil structures significantly extending into the third dimension.  Although the membrane is important here, there is little structural difference to the stiff membrane case [Fig.~\ref{fig:conf_mf1}(h)].
\end{description}

\subsubsection{Structural phases and transitions}

We will shortly summarize similarities with the stiff membrane system here and then look at the differences and emergence of new phenomena in more detail.

\begin{figure}
  \centering
  \includegraphics[width=\linewidth]{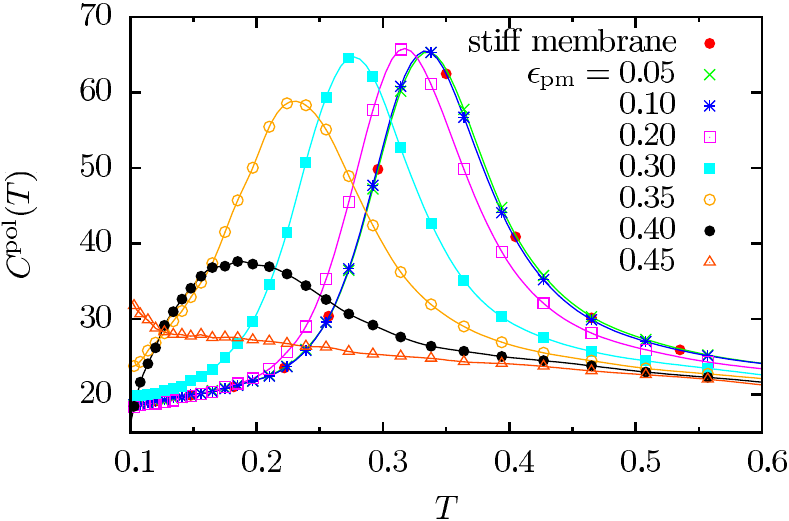}\\
  \includegraphics[width=\linewidth]{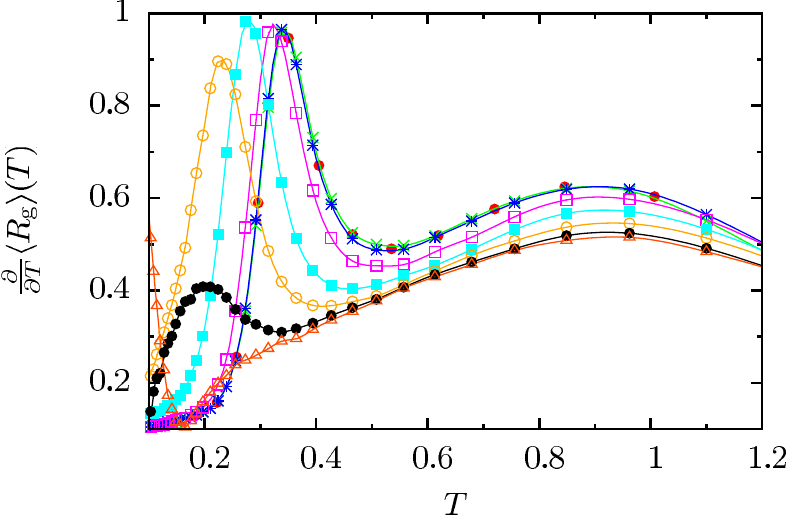}
  \caption[Polymer heat capacity and temperature derivative of the radius of gyration for low polymer-membrane interaction strengths in the flexible membrane system]
  {(Color online) Polymer heat capacity (top) and temperature derivative of the radius of gyration (bottom) for low polymer-membrane interaction strengths in the flexible membrane system.  For better comparison, the values for the stiff membrane system at $\epsilon_\mathrm{pm} = 0.05$ (Fig.~\ref{plot:mf0_dc-g-de}) are plotted again (red circles).  The legend applies to both plots.}
  \label{plot:mf1_dc-g-de}
\end{figure}
The phases \textsf{DC}, \textsf{G}, and \textsf{DE} are hardly affected by the presence of the membrane.  Therefore, we expect the fluctuations of the membrane to have little effect on these phases and the transitions between them.  For the adsorbed random coil phase \textsf{AE2}, the membrane is, of course, essential as the adsorbing surface.  Nevertheless, the phase is already highly disordered such that membrane fluctuations should not generate new effects here.  This is exactly what we observe in our simulation results and is reflected in the phase diagram where we just have marginal differences to the stiff membrane system.
We do not repeat all the plots of the observables identifying the phases just mentioned and their transitions.  To provide only one example for these obvious similarities, Fig.~\ref{plot:mf1_dc-g-de} shows the polymer heat capacity and the temperature derivative of the radius of gyration for low interaction strengths, $0.05 \leq \epsilon_\mathrm{pm} \leq 0.45$.  The curves clearly resemble those in Fig.~\ref{plot:mf0_dc-g-de} for the stiff membrane system indicating, again, the $\Theta$ collapse around \revis{$T \approx 0.9$} and the freezing transition at \revis{$T \approx 0.3$} of the polymer.  Note that we already find a small difference to the stiff membrane system here.  The freezing peak is still pronounced for $\epsilon_\mathrm{pm} = 0.35$ and present up to $\epsilon_\mathrm{pm} = 0.40$ where this transition already disappeared in the stiff membrane system.  This possibly indicates a slightly higher stability of the icosahedral phase in the flexible membrane system.  A reason for this could be the membrane's possibility to adapt to the polymer structure and build up more polymer-membrane contacts while the polymer maintains its icosahedral shape.

\begin{figure}
  \centering
  \includegraphics[width=\linewidth]{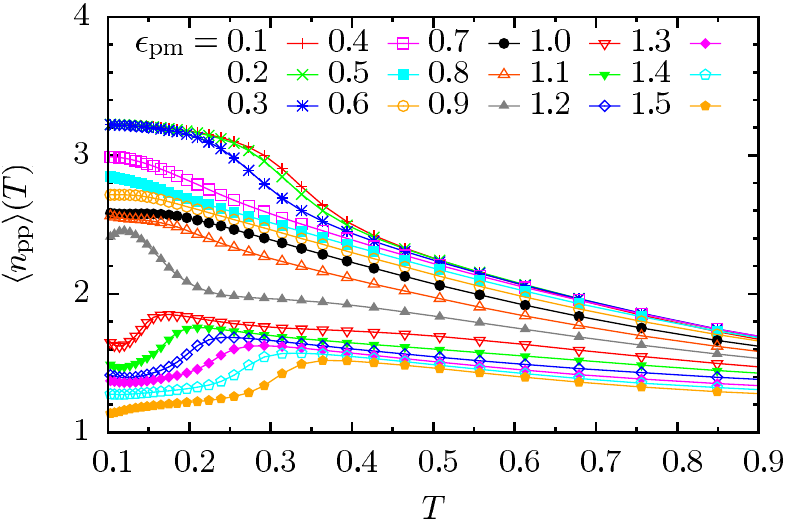}\\
  \includegraphics[width=\linewidth]{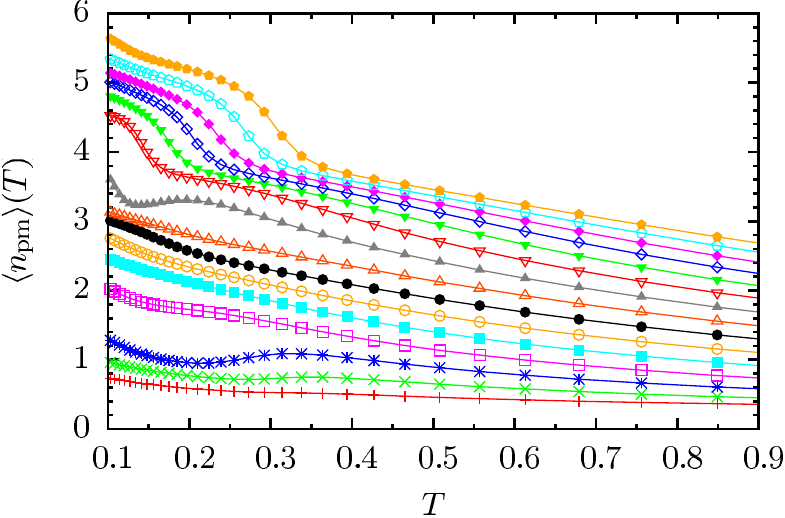}
  \caption[Mean contact numbers in the flexible membrane system]
  {(Color online) Mean number of intrinsic (top) and polymer-membrane (bottom) contacts as a function of temperature for various polymer-membrane interaction strengths in the flexible membrane system.  The legend applies to both plots.}
  \label{plot:mf1_lowT_n}
\end{figure}
In the stiff membrane system, the mean numbers of intrinsic and polymer-membrane contacts showed a striking convergence to our predicted values for the highly ordered low-temperature states.  For the case of a flexible membrane, the curves do not exhibit such a clear convergence (Fig.~\ref{plot:mf1_lowT_n}), but still, we can extract some information about the low-temperature behavior of the system.
For $\epsilon_\mathrm{pm} \leq 0.3$, again we observe convergence to a mean of $3.23$ intrinsic contacts uniquely characterizing the icosahedral phase \textsf{DC}.

For $\epsilon_\mathrm{pm} \geq 1.0$, the mean number of polymer-membrane contacts increases significantly above $4$ while the mean number of intrinsic contacts drops down to values even below those of the disordered random coil phases.  The latter number suggests that the probability of specifically stretched structures is enhanced in this region, whereas in random coils expanded structures only dominate because of higher entropy.  The former number can only be reached when the membrane locally wraps around the individual monomers allowing for more than four polymer-membrane contacts.  In the most perfect realization of a structure in the embedded elongated phase \textsf{ME}, the polymer would be linearly stretched, yielding $12/13 \approx 0.92$ intrinsic contacts, and incorporated in a channel with a triangular cross section such that every monomer is in optimal distance to two square patches of the membrane, resulting in a number of six polymer-membrane contacts per monomer.  Obviously, this idealized situation is not reached here but from the tendencies shown in Fig.~\ref{plot:mf1_lowT_n} we propose embedded elongated conformations (\textsf{ME}) as predominant states for low temperature and high polymer-membrane interaction strength.  In the range of $0.7 \leq \epsilon_\mathrm{pm} \leq 0.8$ the two contact numbers seem to converge to certain intermediate values.  Although we are not sure which would be the perfect symmetry in this case, we expect to find the embedded compact phase \textsf{MC} as the low-temperature phase here as suggested by the snapshot in Fig.~\ref{fig:conf_mf1}(e).

\begin{figure}
  \centering
  \includegraphics[width=\linewidth]{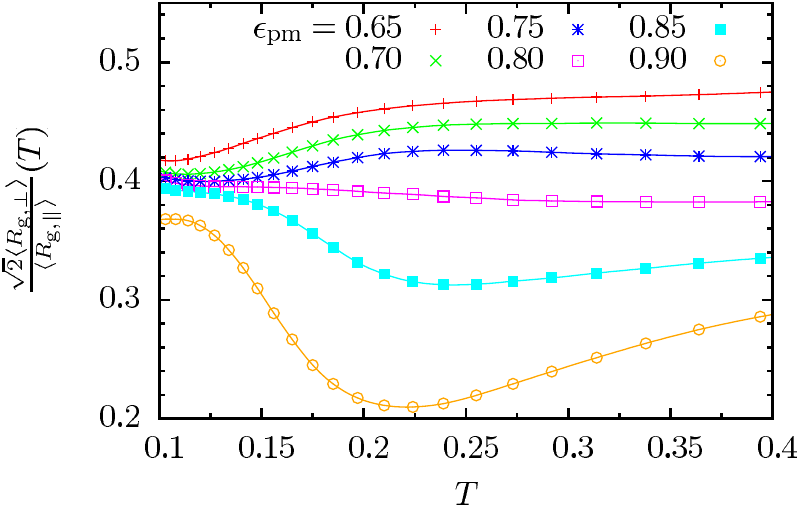}
  \caption[Sphericity aspect ratio at the \textsf{MC}~--~\textsf{G} and \textsf{MC}~--~\textsf{AC1} transitions in the flexible membrane system]
  {(Color online) Sphericity aspect ratio at the \textsf{MC}~--~\textsf{G} and \textsf{MC}~--~\textsf{AC1} transitions in the flexible membrane system.}
  \label{plot:mf1_mc-g_rgrat}
\end{figure}
\begin{figure}
  \centering
  \includegraphics[width=\linewidth]{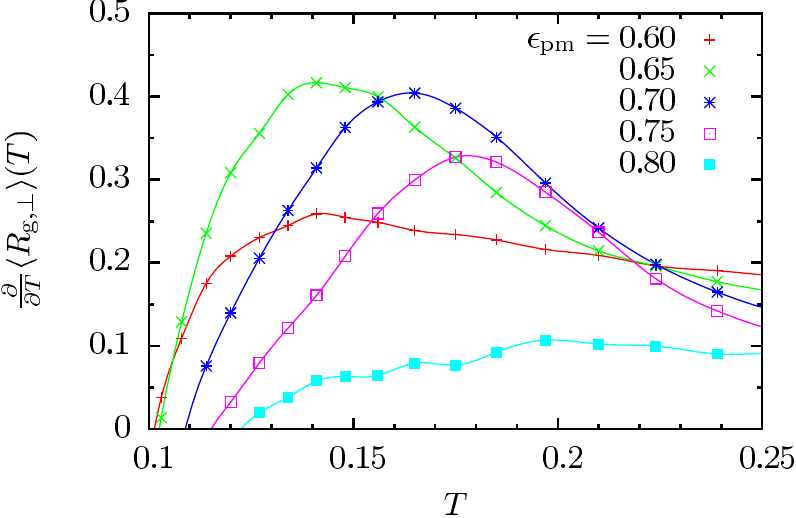}
  \caption[Temperature derivative of the $z$ component of the gyration tensor at the \textsf{MC}~--~\textsf{G} transition in the flexible membrane system]
  {(Color online) Temperature derivative of the perpendicular component of the gyration tensor at the \textsf{MC}~--~\textsf{G} transition in the flexible membrane system.}
  \label{plot:mf1_mc-g_dtrgz}
\end{figure}
\begin{figure}
  \centering
  \includegraphics[width=\linewidth]{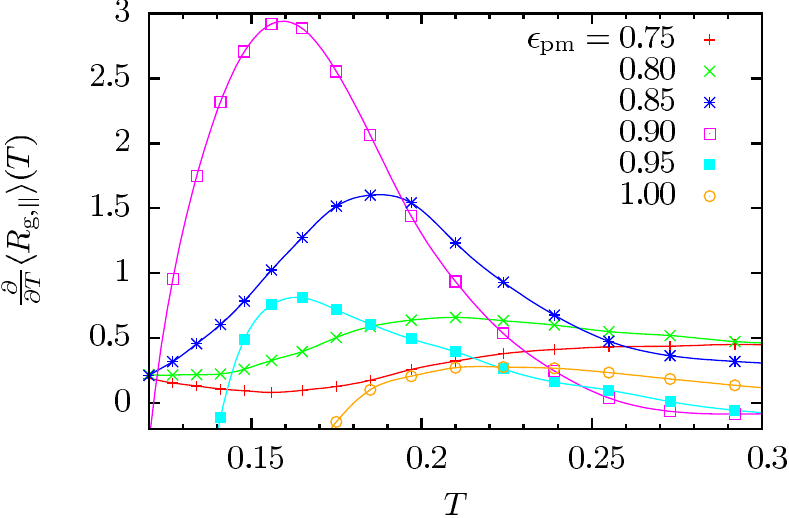}
  \caption[Temperature derivative of the $xy$ component of the gyration tensor at the \textsf{MC}~--~\textsf{AC1} transition in the flexible membrane system]
  {(Color online) Temperature derivative of the parallel component of the gyration tensor at the \textsf{MC}~--~\textsf{AC1} transition in the flexible membrane system.}
  \label{plot:mf1_mc-ac1_dtrgxy}
\end{figure}

The transition out of the \textsf{MC} phase with increasing temperature is analyzed in Figs.~\ref{plot:mf1_mc-g_rgrat}--\ref{plot:mf1_mc-ac1_dtrgxy}.  In the range of $0.65 \lesssim \epsilon_\mathrm{pm} \lesssim 0.75$, the perpendicular component of the gyration tensor significantly increases in the region $0.14 \lesssim T \lesssim 0.18$ as indicated by the peaks in the corresponding temperature derivative (Fig.~\ref{plot:mf1_mc-g_dtrgz}).  Additionally, we observe a slight increase in the sphericity aspect ratio as depicted in Fig.~\ref{plot:mf1_mc-g_rgrat} meaning that conformations change from oblate to more spherical shapes.  This marks the transition from embedded compact (\textsf{MC}) to globular (\textsf{G}) conformations which qualitatively compares to the droplet -- globule transition (\textsf{AC2}~--~\textsf{G}) in the stiff membrane case.

For higher polymer-membrane interaction strengths, $0.80 \lesssim \epsilon_\mathrm{pm} \lesssim 0.95$, we observe a peak in the temperature derivative of the parallel component of the gyration tensor in the region $0.15 \lesssim T \lesssim 0.19$ (Fig.~\ref{plot:mf1_mc-ac1_dtrgxy}).  As the perpendicular component does not increase simultaneously, this indicates a further ``flattening'' of the polymer structures.  This can also be concluded from the drop in the sphericity aspect ratio (Fig.~\ref{plot:mf1_mc-g_rgrat}).  Together with the observed drop in the number of intrinsic contacts (Fig.~\ref{plot:mf1_lowT_n}), we identify the transition from the embedded compact (\textsf{MC}) to the adsorbed compact (\textsf{AC1}) phase here.

\begin{figure}
  \centering
  \includegraphics[width=\linewidth]{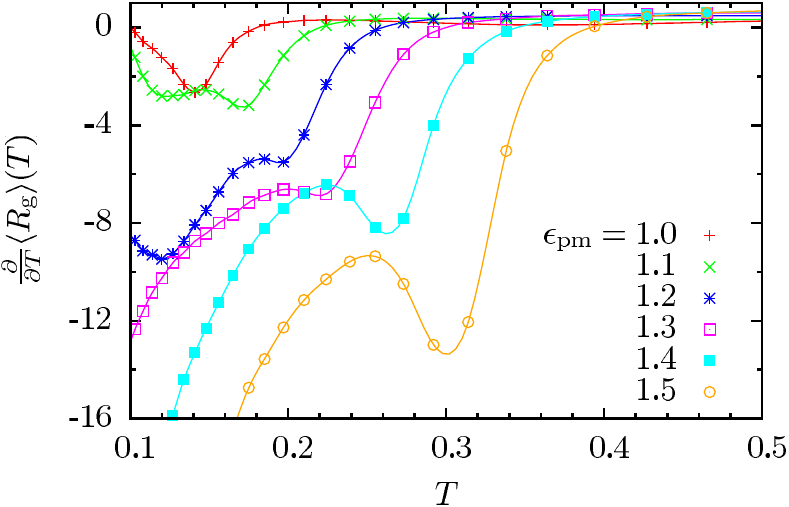}\\
  \includegraphics[width=\linewidth]{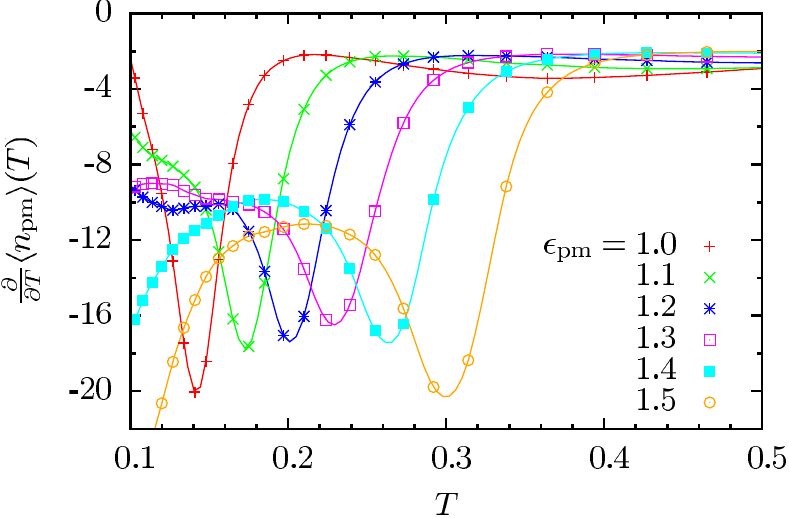}
  \caption[Temperature derivatives of the radius of gyration and the mean number of polymer-membrane contacts at the \textsf{ME}~--~\textsf{AC1} transition in the flexible membrane system]
  {(Color online) Temperature derivatives of the radius of gyration (top) and the mean number of polymer-membrane contacts (bottom) at the \textsf{ME}~--~\textsf{AC1} transition in the flexible membrane system.}
  \label{plot:mf1_me-ac1}
\end{figure}
One of the strongest transition signals we find is shown in Fig.~\ref{plot:mf1_me-ac1}.  The temperature derivatives of the radius of gyration and the number of polymer-membrane contacts exhibit a deep valley indicating a sharp transition line.  The position of the valley increases from $T \approx 0.14$ at $\epsilon_\mathrm{pm} = 1.0$ to $T \approx 0.3$ at $\epsilon_\mathrm{pm} = 1.5$ almost linearly.  
The decrease in the radius of gyration suggests that we have a transition from expanded (or elongated) structures at lower temperature to more compact structures at higher temperature.  This is exactly the opposite of what we usually observe in transitions like the $\Theta$ collapse and would not be possible without the membrane serving as an agent or a medium enhancing the probability of elongated conformations in the low-temperature phase.  
Looking also at the mean number of polymer-membrane contacts per monomer (Fig.~\ref{plot:mf1_lowT_n}), we see that the ``inverse collapse'' is accompanied by a drop from more than four to slightly less than four contacts per monomer, which was the maximum value for a stiff membrane.  This suggests that the membrane adapts its shape to the polymer to form up to six contacts with each monomer in the low-temperature phase and gives up this behavior at higher temperatures after passing the transition.
Taken together, we observe a transition form elongated conformations of the polymer which are partially incorporated into the membrane (\textsf{ME}) to disk-like compact film structures adsorbed to the fluctuating membrane surface (\textsf{AC1}).

\begin{figure}
  \centering
  \includegraphics[width=\linewidth]{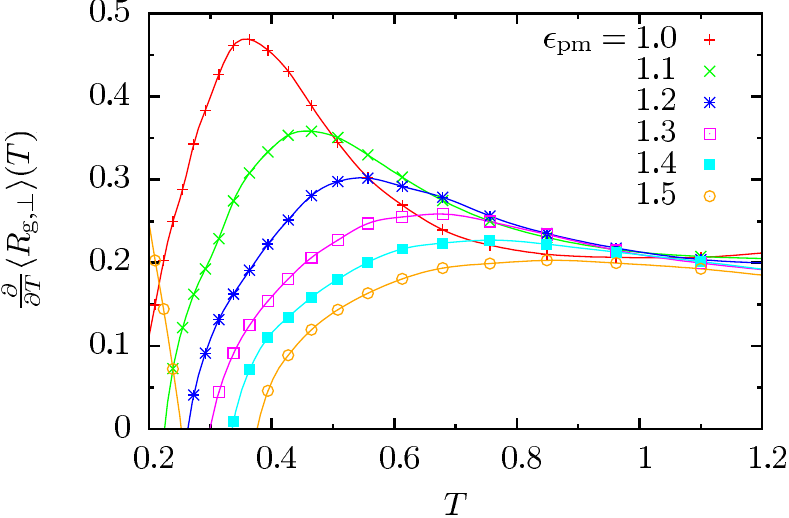}\\
  \includegraphics[width=\linewidth]{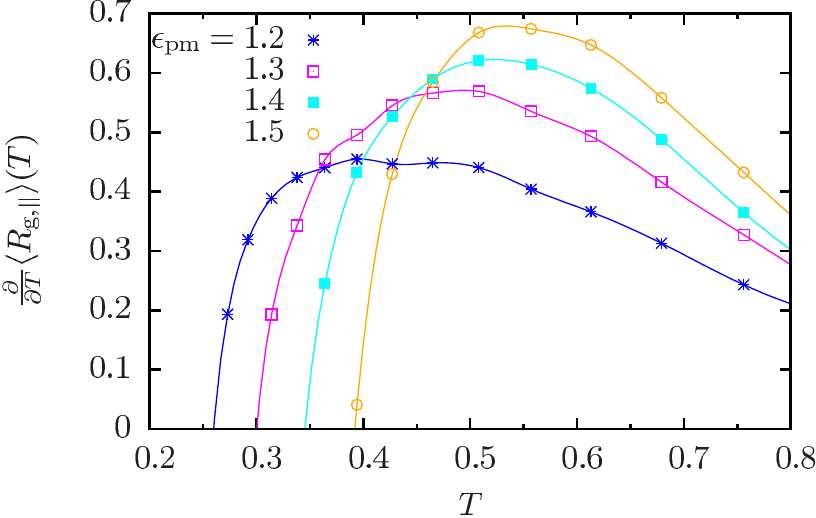}
  \caption[Temperature derivative of the components of the gyration tensor at the \textsf{AC1}~--~\textsf{AE2} / \textsf{AE1} transition in the flexible membrane system]
  {(Color online) Temperature derivative of the perpendicular (top) and parallel (bottom) components of the gyration tensor at the \textsf{AC1}~--~\textsf{AE2} / \textsf{AE1} transition in the flexible membrane system.}
  \label{plot:mf1_ac1-ae}
\end{figure}
For higher temperatures we observe another, but much weaker, transition signal in the temperature derivative of the radius of gyration.  Taking a closer look at the temperature derivatives of the individual components of the gyration tensor (Fig.~\ref{plot:mf1_ac1-ae}), we find a similar behavior to what we have seen already in the stiff membrane system.  The perpendicular component increases at a transition line ranging from $T \approx 0.35$ at $\epsilon_\mathrm{pm} = 1.0$ to $T \approx 0.85$ at $\epsilon_\mathrm{pm} = 1.5$ indicating a (partial) desorption of the polymer from the membrane surface.  Additionally, the parallel component increases along a transition line from $T \approx 0.4$ at $\epsilon_\mathrm{pm} = 1.2$ to $T \approx 0.55$ at $\epsilon_\mathrm{pm} = 1.5$, indicating an expansion of the polymer parallel to the membrane surface.  As in the stiff membrane case, we conclude that what we observe here is the transition from disk-like compact film structures adsorbed to the membrane surface (\textsf{AC1}) to expanded random coil structures, partially adsorbed to the membrane surface (\textsf{AE2}).  For high polymer-membrane interaction strengths, we think that additionally (almost) completely adsorbed random coil conformations (\textsf{AE1}) appear in between.

In the region of the phase diagram where the phases \textsf{G}, \textsf{MC}, \textsf{AC1}, and \textsf{AE2} meet, $\epsilon_\mathrm{pm} \approx 0.7, \dots, 1.0$ and $T \approx 0.2, \dots, 0.6$, many transitions seem to superimpose.  Therefore, it cannot be clearly decided what structures are predominantly present in that region.  As this is a finite-size effect, we think that investigations on larger systems could reveal more about the actual phase structure here.

\section{\label{sec:summary}Summary}
In this paper we have studied the structural behavior of a coarse-grained model system consisting of a single elastic, flexible polymer in contact with a flexible membrane.
The main goal was to identify the conformational thermodynamic pseudophases of the system in dependence of temperature and polymer-membrane interaction strength as external parameters.  To clearly point out the effects emerging from the flexibility and adaptivity of the membrane, we separately investigated the two cases of a stiff membrane forced into a flat state and a flexible membrane, both with the polymer grafted.

The pseudophase diagrams were constructed from the measured observables and their fluctuations for an intermediate-sized system of polymer length $13$ and membrane size $27 \times 27$.  The observed conformational pseudophases include the following:
\begin{itemize}
\item Desorbed conformations for low polymer-membrane interaction strengths.  These phases are well-known from studies of free polymers.  With decreasing temperature the polymer undergoes two transitions, the $\Theta$ transition from extended random coil structures (\textsf{DE}) to more compact but disordered globular conformations (\textsf{G}) and the freezing transition to a highly compact icosahedral ground state (\textsf{DC}).
\item Adsorbed disordered conformations in both systems.  In these conformations the polymer is adsorbed to the membrane surface but still highly disordered such that the presence or absence of membrane fluctuations does not induce qualitative differences.  We distinguish between (almost) completely adsorbed, extended (\textsf{AE1}) and partially adsorbed, extended (\textsf{AE2}) random coil structures.
\item Adsorbed well-ordered conformations in the stiff membrane system.  For low temperatures, we observed highly ordered conformations adsorbed to the membrane surface.  We distinguish between disk-shaped compact film structures (\textsf{AC1}) and semi-spherical droplets (\textsf{AC2}).
\item Embedded conformations in the flexible membrane system.  At low temperatures, we found the membrane adapting its structure such that it partially incorporates the polymer.  Compact oblate-shaped structures embedded into the membrane (\textsf{MC}) were observed at intermediate and linearly stretched embedded structures (\textsf{ME}) were identified at high interaction strength.
\end{itemize}

\revis{Because of its smallness, the system's structural behavior is influenced
  by finite-size effects. Nevertheless, we expect that the phase behavior of the
system remains qualitatively intact also for larger systems. A verification by a
systematic finite-size analysis is currently unachievable for most of the
phases, in particular in the low-temperature regime. Thus, this remains future
work in its own right. However, also the finite-size effects are of substantial
interest since classes of polymers, such as proteins, are naturally finite.}
\section{Acknowledgments}
This work is partially supported by grant No. JA483/24-3 of the
   Deutsche Forschungsgemeinschaft (DFG) and 
by a supercomputer time grant of the John von Neumann Institute for Computing (NIC), Forschungszentrum J\"ulich, under No.\ jiff39.

\begin{thebibliography}{99}
%
\bibitem{hegger_chain_1994}
R. Hegger and P. Grassberger, J. Phys. A \textbf{27}, 4069 (1994).
%
\bibitem{vrbova_adsorption_1996}
T. Vrbov\'a and S. G. Whittington, J. Phys. A \textbf{29}, 6253 (1996).
%
\bibitem{vrbova_adsorption_1998}
T. Vrbov\'a and S. G. Whittington, J. Phys. A \textbf{31}, 3989 (1998).
%
\bibitem{vrbova_adsorption_1999}
T. Vrbov\'a and K. Prochzka, J. Phys. A \textbf{32}, 5469 (1999).
%
\bibitem{singh_crossover_2001}
Y. Singh, D. Giri, and S. Kumar, J. Phys. A \textbf{34}, L67 (2001).
%
\bibitem{rajesh_adsorption_2002}
R. Rajesh, D. Dhar, D. Giri, S. Kumar, and Y. Singh, Phys. Rev. E \textbf{65}, 056124 (2002).
%
\bibitem{krawczyk_layering_2005}
J. Krawczyk, A. L. Owczarek, T. Prellberg, and A. Rechnitzer, Europhys. Lett. \textbf{70}, 726 (2005).
%
\bibitem{bachmann_conformational_2005}
M. Bachmann and W. Janke, Phys. Rev. Lett. \textbf{95}, 058102 (2005).
%
\bibitem{bachmann_substrate_2006}
M. Bachmann and W. Janke, Phys. Rev. E \textbf{73}, 041802 (2006).
%
\bibitem{kallrot_dynamic_2007}
N. K\"allrot and P. Linse, Macromolecules \textbf{40}, 4669 (2007).
%
\bibitem{moeddel_conformational_2009}
M. M\"oddel, M. Bachmann, and W. Janke, J. Phys. Chem. B \textbf{113}, 3314 (2009).
%
\bibitem{moeddel_systematic_2010}
M. M\"oddel, W. Janke, and M. Bachmann, Phys. Chem. Chem. Phys. \textbf{12}, 11548 (2010).
%
\bibitem{breidenich_shape_2000}
M. Breidenich, R. R. Netz, and R. Lipowsky, Europhys. Lett. \textbf{49}, 431 (2000).
%
\bibitem{breidenich_adsorption_2001}
M. Breidenich, R. R. Netz, and R. Lipowsky, Eur. Phys. J. E \textbf{5}, 403 (2001).
%
\bibitem{auth_self-avoiding_2003}
T. Auth and G. Gompper, Phys. Rev. E \textbf{68}, 051801 (2003).
%
\bibitem{auth_fluctuation_2005}
T. Auth and G. Gompper, Phys. Rev. E \textbf{72}, 031904 (2005).
%
\bibitem{grest_molecular_1986}
G. S. Grest and K. Kremer, Phys. Rev. A \textbf{33}, 3628 (1986).
%
\bibitem{schnabel_surface_2009}
S. Schnabel, T. Vogel, M. Bachmann, and W. Janke, Chem. Phys. Lett. \textbf{476}, 201 (2009).
%
\bibitem{schnabel_elastic_2009}
S. Schnabel, M. Bachmann, and W. Janke, J. Chem. Phys. \textbf{131}, 124904 (2009).
%
\bibitem{seaton_developments_2008}
D. T. Seaton, S. J. Mitchell, and D. P. Landau, Braz. J. Phys. \textbf{38}, 48 (2008).
%
\bibitem{seaton_wang_2009}
D. T. Seaton, T. W\"ust, and D. P. Landau, Comput. Phys. Commun. \textbf{180}, 587 (2009).
%
%
\revis{
\bibitem{rem1}
Although it would be more appropriate to
    call it an oligomer rather
    than a polymer, we keep the latter term since we think that our main results
  are also valid for larger systems. Furthermore, the term ``oligomer'' is
  frequently used in connection with multiple-chain systems and thus might lead to
  confusion here.
}
%
\bibitem{bird_dynamics_1987}
R. B. Bird, C. F. Curtiss, R. C. Armstrong, and O. Hassager, \textit{Dynamics of Polymeric Liquids}, 2nd ed., (Wiley, New York, 1987), Vol. 2.
%
\bibitem{milchev_formation_2001}
A. Milchev, A. Bhattacharya, and K. Binder, Macromolecules \textbf{34}, 1881 (2001).
%
\bibitem{kantor_statistical_1986}
Y. Kantor, M. Kardar, and D. R. Nelson, Phys. Rev. Lett. \textbf{57}, 791 (1986).
%
\bibitem{popova_structure_2007}
H. Popova and A. Milchev, J. Chem. Phys. \textbf{127}, 194903 (2007).
%
\bibitem{popova_adsorption_2008}
H. Popova and A. Milchev, J. Chem. Phys. \textbf{129}, 215103 (2008).
%
\bibitem{geyer_annealing_1995}
C. J. Geyer and E. A. Thompson, J. Am. Stat. Assoc. \textbf{90}, 909 (1995).
%
\bibitem{hukushima_exchange_1996}
K. Hukushima and K. Nemoto, J. Phys. Soc. Jpn. \textbf{65}, 1604 (1996).
%
\bibitem{metropolis_equation_1953}
N. Metropolis, A.~W. Rosenbluth, M.~N. Rosenbluth, A.~H. Teller, and E. Teller, J. Chem. Phys. \textbf{21}, 1087 (1953).
%
\bibitem{bittner_make_2008}
E. Bittner, A. Nu\ss{}baumer, and W. Janke, Phys. Rev. Lett. \textbf{101}, 130603 (2008).
%
\bibitem{ferrenberg_optimized_1989}
A. M. Ferrenberg and R. H. Swendsen, Phys. Rev. Lett. \textbf{63}, 1195 (1989).
%
\end{thebibliography}
\end{document}